\documentclass[titlepage]{article}
\title{\vspace{-3cm}}

\usepackage{authblk} % For nice title page
\usepackage[utf8]{inputenc}
\usepackage[margin=1in]{geometry}
\usepackage[titletoc,title]{appendix}
\usepackage[svgnames]{xcolor}
\usepackage{listings}
\usepackage{bm}
\usepackage{graphicx,float}

\usepackage{float}
\usepackage[para,online,flushleft]{threeparttable}
\usepackage{booktabs}% http://ctan.org/pkg/booktabs
\newcommand{\tabitem}{~~\llap{\textbullet}~~}

\usepackage{setspace}
\graphicspath{{Images/}}

\vspace{-8ex}
\date{}

\title{\textbf{Statistical analyses of ordinal outcomes in randomised controlled trials: protocol for a scoping review}} 
\author[1,2,*]{\normalsize{Chris J. Selman}}
\author[1,2]{Katherine J. Lee}
\author[1,3]{Robert K. Mahar}

\affil[1]{Clinical Epidemiology and Biostatistics Unit, Murdoch Children's Research Institute, Parkville, Victoria 3052, Australia.}

\affil[2]{Department of Paediatrics, University of Melbourne, Parkville, Victoria 3052, Australia.} 

\affil[3]{Centre for Epidemiology and Biostatistics, Melbourne School of Population and Global Health, University of Melbourne, Parkville, Victoria 3052, Australia.}

\affil[*]{Corresponding author: Chris Selman, Clinical Epidemmiology and Biostatistics Unit, Murdoch Children's Research Institute, Parkville, Victoria 3052, Australia. chris.selman@mcri.edu.au}

\date{\today}

\begin{document}
\maketitle

\newpage

\section{ABSTRACT}

\textbf{Introduction}
Randomised controlled trials aim to assess the impact of one (or more) health interventions relative to other standard interventions. RCTs sometimes use an ordinal outcome, which is an endpoint that comprises of multiple, monotonically ordered categories that are not necessarily separated by a quantifiable distance. Ordinal outcomes are appealing in clinical settings as specific disease states can represent meaningful categories that may be of clinical importance to researchers. They can also retain information and increase statistical power compared to dichotomised outcomes, and can allow multiple clinical outcomes to be comprised in a single endpoint. Target parameters for ordinal outcomes in RCTs may vary depending on the nature of the research question, the modelling assumptions, and the expertise of the data analyst.

The aim of this scoping review is to systematically describe the use of ordinal outcomes in contemporary RCTs. Specifically, we aim to:
\begin{itemize}
    \item Identify which target parameters are of interest in trials that use an ordinal outcome, and whether these are explicitly defined.
    \item Describe how ordinal outcomes are analysed in RCTs to estimate a treatment effect.
    \item Describe whether RCTs that use an ordinal outcome adequately report key methodological aspects specific to the analysis of the ordinal outcome.
\end{itemize} Results from this review will outline the current state of practice of the use of ordinal outcomes in RCTs. Ways to improve the analysis and reporting of ordinal outcomes in RCTs will be discussed. 

\textbf{Methods and Analysis}
We will review RCTs that are published in the top four medical journals (British Medical Journal, New England Journal of Medicine, The Lancet and the Journal of the American Medical Association) between 1 January 2012 and 31 July 2022 that use an ordinal outcome as a primary or secondary outcome. The review will be conducted using PubMed. Our review will adhere to guidelines for scoping reviews as described in the PRISMA-ScR checklist. The study characteristics and details of the study design and analysis, including the target parameter(s) and statistical methods used to analyse the ordinal outcome, will be extracted from eligible studies. The screening, review, and data extraction will be conducted using Covidence, a web-based tool for managing systematic reviews. The data will be summarised using descriptive statistics.

\newpage 
\section{INTRODUCTION}
\subsection{Background}
Randomised controlled trials (RCTs) typically aim to estimate the average causal effect of one intervention relative to at least one other intervention. The average causal effect can take many forms, such as a risk or odds ratio (binary outcome) or difference in means (continuous outcome). The outcome of interest in a trial may be nominal, ordinal, interval or a ratio \cite{stevens1946theory}. Nominal outcomes refer to outcomes that are categorical and unranked, such as the blood type of a patient. If the outcome is measured on an interval scale, the outcome can be categorised, ranked, and the difference between any two proximate values are equally spaced (e.g. body temperature). In addition to these properties, the ratio scale has a true zero point (e.g. weight). Ordinal scales can be considered to inhabit the space between nominal and interval/ratio scales; they are categorised and ranked, but the distance between any two categories is not necessarily meaningfully quantifiable or equally spaced. For example, a change from a disease-free state to hospitalisation would not be considered to be equivalent to a change from hospitalisation to death. An ordinal outcome is therefore a variable that measures a single characteristic that comprises of multiple, monotonically ordered categories that are not separated by a meaningful distance \cite{velleman1993nominal}. The categories should also be mutually exclusive (such that the categories are non-overlapping), detect improvement and deterioration, and be unambiguously defined (so that categories can be clearly distinguished from each other) \cite{mackenzie1986standards}. An additional condition of an ordinal outcome, if it measures a change between two points in time, is that the scale should be symmetrical in structure to avoid bias \cite{mackenzie1986standards}. That is, there should be an equal number of categories that represent both improvement and deterioration. 

Ordinal outcomes have become increasingly common in trials, particularly during the COVID-19 pandemic. For example, the World Health Organisation (WHO) developed the WHO Clinical Progression Scale \cite{marshall2020minimal}, an ordinal scale that describes disease severity of COVID-19 that has been adapted in various treatment trials \cite{lovre2021acute, song2022treatment}. The categories reflect patient states that include being uninfected with COVID-19, ambulatory mild disease (asymptomatic or symptomatic), moderate disease (defined by patient hospitalisation and whether oxygen therapy is required or not), severe disease (ranging from a hospitalised patient who requires oxygen by non-invasive ventilation or high flow, to patients requiring mechanical ventilation with any of vasopressors, dialysis or extracorporeal membrane oxygenation for treatment), and death \cite{marshall2020minimal}. A review of clinical trials on the management of COVID-19 found that over half of the trials that evaluated disease severity and progression used an ordinal outcome \cite{mathioudakis2020outcomes}, where the majority were used as secondary outcomes. Ordinal outcomes are also commonly used in stroke trials that often use the modified Rankin scale, a measure of the degree of disability among individuals who have suffered a stroke, as an outcome of interest \cite{banks2007outcomes, de2022effect, hubert2022association, bosel2022effect}. 

Although it can be easier to interpret a clinically important effect using a dichotomised or continuous outcome, various disease states measured on an ordinal scale can represent meaningful distinctions that may be of clinical importance. Ordinal outcomes are also appealing as they can retain information and increase statistical power compared to dichotomised outcomes, allowing a smaller sample size to be used \cite{roozenbeek2011added}. They can also answer important clinical questions regarding specific patient states that cannot be answered using continuous outcomes, and can allow multiple clinical outcomes to be comprised in a single endpoint. Although there are advantages to using ordinal outcomes, the required analyses can be complicated and important considerations need to be made in the design phase of the trial. For example, the number of categories in the ordinal scale (fewer categories may reduce power and increase the sample size needed to detect an effect \cite{peterson2017analysis}), and explicitly defining an appropriate target parameter to compare the intervention groups.

There are a number of different target parameters that may be used to compare interventions with an ordinal outcome. For example, one could use an odds ratio that is assumed to be constant across all of the dichotomisations of the ordinal scale, known as the proportional odds assumption. Such a statistic can be estimated using the cumulative logit model, commonly referred to as the `proportional odds model' \cite{mccullagh1980regression}. Alternatively, the target parameters of interest might be odds ratios that use a baseline category as the reference level that can be analysed using multinomial regression. Ordinal outcomes can also be dichotomised for analysis, with the target parameter of interest possibly being the difference in proportions between the intervention groups which can be estimated using binomial regression. Finally, the ordinal outcome can be treated as continuous data with the target parameter of interest being a difference in means, estimated using a linear regression model. 

There are pros and cons to the different target parameters and methods of analysis for ordinal outcomes. When an ordinal outcome has been dichotomised for analysis, the analyses are simple and the interpretation of the effect of interest can be easily understood. Ordinal outcomes that have been dichotomised, however, can discard potentially useful information on the levels of the scale and can lower statistical power compared to the original ordinal outcome \cite{altman2006cost, d2020ordinal, dodd2020endpoints}. Armstrong and Sloan found that the variance of the odds ratio using a logistic regression model is between 25--50\% higher than the variance of the odds ratio estimated from a cumulative logit model \cite{armstrong1989ordinal}, therefore reducing the power to detect a clinically important effect. When an ordinal outcome is treated as a continuous outcome, a linear regression model can be used to estimate a difference in means. Such an analysis assumes that the difference between any two categories of the ordinal outcome is uniform and separated by a quantifiable distance, the outcome has unbounded support, and that the outcome follows a normal distribution. Although the analyses are straightforward, these assumptions are likely to be violated as ordinal outcomes often have few categories that is insufficient to approximate a normal distribution and, more importantly, the distance between any two categories can not be described quantitatively. Any treatment effect estimated from such an analysis would therefore be difficult to interpret. 

If the target parameter is an odds ratio from a proportional odds model, then the target parameter has a fairly simple clinical interpretation (usually being the odds of a better outcome). However, in practice, the proportional odds assumption may not hold \cite{kim2003assessing, fullerton2012proportional}, in which case the odds ratios across each binary split of the ordinal outcome are not equal. Instead, the treatment comparison can be extended to be odds ratios that are not constant across the binary splits of the ordinal scale, which can be estimated using a partial proportional odds model \cite{peterson1990partial}. Alternatively, adjacent-category logit and continuation ratio models could also be used to estimate the odds ratios, though these models have different model assumptions and interpretations of the target parameter(s). All these models also have natural extensions to account for repeated measures over time (e.g. mixed models \cite{agresti1993proportional} or Markov transition models \cite{de2017proportional}, \cite{mhoon2010continuous}). 

There has been some methodological research that describe how ordinal outcomes can be used in in specific settings, such as vascular prevention trials \cite{bath2008use} and comparative studies \cite{scott1997statistical}. However, these studies focussed on a small number of statistical models and are not reflective of more general settings. With the increasing use of ordinal outcomes in randomised trials should come an improved understanding of how ordinal outcomes are used in practice. Better understanding will ensure that any issues in the use of ordinal outcomes in RCTs are identified and improvements to the reporting and analyses of such outcomes can be discussed. We aim to improve understanding by conducting a scoping review to systematically review the literature to \textit{(i)} identify which target parameters are of interest in trials that use an ordinal outcome and whether these are explicitly defined; \textit{(ii)} describe how ordinal outcomes are analysed in RCTs to estimate a treatment effect; and \textit{(iii)} describe whether RCTs that use an ordinal outcome adequately report key methodological aspects specific to the analysis of the ordinal outcome.

\newpage
\section{METHODS AND ANALYSIS}
\subsection{Search strategy}
We anticipate that the expected start date of this review is 15 August 2022 and the anticipated completion date is 1 February 2023. We will systematically search, identify, and describe RCTs that have used an ordinal outcome that have been published in the top four medical journals between 1 January 2012 and 31 July 2022. The four medical journals that will be included in this search are \textit{British Medical Journal} (BMJ), \textit{New England Journal of Medicine} (NEJM), \textit{The Lancet}, and the \textit{Journal of the American Medical Association} (JAMA). These journals have been selected because they are the top journals in the medical field that publish original, peer-reviewed research from RCTs and have been used in other reviews of trials \cite{bell2014handling, berwanger2009quality}. It is expected that these journals will capture the current best practice in the use of ordinal outcomes in RCTs.

We will identify RCTs to be included in the review by searching PubMed. Our search terms will employ search strategies developed to identify RCTs \cite{higgins2011cochrane}, and terms that are used to describe ordinal outcomes in the title and abstract of relevant published articles. Since we anticipate that varied terminologies are used to describe ordinal outcomes, we first examined various RCTs that use an ordinal outcome to determine the type of terminology that is used to describe ordinal outcomes. This enabled us to develop and refine our search strategy. The full search strategy to be used in the review is outlined in Table \ref{tab:table1}.

\renewcommand{\arraystretch}{1.6}
\begin{table}[H]
\begin{center}
\begin{threeparttable}
\caption{PubMed Search Strategy}
\centering
\begin{tabular}{|l|}
\hline
\multicolumn{1}{|c|}{\textbf{Search strategy}} \\ \hline
\begin{tabular}[c]{@{}l@{}}(JAMA{[}journal{]}  OR NEJM{[}journal{]} OR lancet{[}journal{]} OR BMJ{[}journal{]}) AND \\ (ordinal{[}tiab{]\tnote{1} } OR   categorical{[}tiab{]} OR multinomial{[}tiab{]} OR “item-response”{[}tiab{]}   \\ OR psychometric{[}tiab{]} OR scale{[}tiab{]} OR Likert{[}tiab{]}) AND (randomized controlled trial{[}pt{]\tnote{2} }\\ OR controlled clinical trial{[}pt{]} OR trial{[}tiab{]} OR randomized{[}tiab{]} OR placebo{[}tiab{]}\\ OR  clinical trials as topic{[}mesh: noexp{]\tnote{3} } OR randomly{[}tiab{]})\end{tabular} \\ \hline
\end{tabular}
\label{tab:table1}
\begin{tablenotes}
\item[1] Indicates that the search is conducted on article titles and abstracts only; \item[2] Corresponds to a publication type to indicate the article's type of information conveyed; \item[3] Corresponds to a medical subject headings such that the explosion feature has been turned off (explosion searches the more specific terms beneath that heading).
\end{tablenotes}
\end{threeparttable}
\end{center}
\end{table}

\subsection{Eligibility Criteria}

\subsubsection{Inclusion Criteria}
The review will include studies that meet the following criteria:
\begin{enumerate}
    \item The study includes at least one RCT.  \\ 
    We will use the Cochrane definition of a RCT, which are studies in which one of two (or more) health interventions are prospectively assigned to individuals using a random/quasi-random method of allocation \cite{higgins2011cochrane}. 
    \item The study was published in one of the top four medical journals between 1 January 2012 and 31 July 2022: \textit{British Medical Journal}, \textit{New England Journal of Medical}, \textit{Journal of the American Medical Association} or \textit{The Lancet}. \\
    For articles with more than one publication date (such as early-view/online publication), only one publication date is required to be between 1 January 2012 and 31 July 2022. If two or more publication dates are between these dates, the earlier date will be recorded.
    \item The study reports an analysis of a primary or secondary ordinal outcome.\\
    Our review will focus on any ordinal outcomes that are used, whether they were specified as a primary or secondary outcome. The ordinal outcome must have multiple, monotonically ordered categories that are not necessarily separated by a quantifiable distance and do not have equal spacing between categories. 

\end{enumerate}

\subsubsection{Exclusion Criteria}
We will exclude studies that meet the following criteria:
\begin{enumerate}
    \item The study is written in a language other than English. \\
    This criteria has been included as we are not capable of translating studies written in other languages.

    \item The study is a methodological paper examining data from an RCT. \\
    This criterion is included because we are only interested in how ordinal outcomes have been used in real-world RCTs. Methodological papers tend to provide motivating examples that may not be representative of RCTs that use an ordinal outcome in practice. 

    \item The study does not provide either an abstract or full-text. 

    \item The study analyses data from non-human subjects only. 

    \item The manuscript provides a commentary, review, opinion or description only. 
    
    \item The manuscript is a protocol or statistical analysis plan. \\
    These manuscripts will be excluded from the review since one of the aims of this review is about what statistical models were reported in the analysis, and whether they have checked and justified the model assumptions.

    \item The only ordinal outcome(s) is(are) measured on an interval scale. \\
    Studies will be excluded if the ordinal outcome is a numeric scale in which differences between proximate values are separated by a quantifiable and equal distance (e.g. the visual analogue scale). Similarly, studies will be excluded if the outcomes were derived from multiple-items measured on an ordinal scale in which the summary variable is also interval data, such as the Hamilton Depression Rating Scale. Outcomes that are inherently interval data can be analysed using conventional and valid statistical methods, such as linear regression. The focus of this review is how ordinal outcomes, whose categories are not equally spaced and any meaningful distance between categories can only be described qualitatively, can be appropriately analysed in RCTs. 

    \item The study is a systematic review and/or provides a meta-analyses of RCTs.
\end{enumerate}

\subsection{Sample Size}
There is no pre-defined sample size. We plan to include all eligible studies that appear in our search using our pre-defined search strategy.

\subsection{Study selection}
Titles and abstracts identified by the search strategy will be extracted into Covidence, a web-based tool for managing systematic reviews \cite{covidence}. The review began with a piloting phase, where two authors (CS, RM) independently assessed 20 abstracts to ensure that the application of the inclusion and exclusion criteria was consistent between reviewers. If there was more than minor disagreement, then the criteria were further refined following discussion between the three reviewers (CS, RM, KL).

The review will be conducted by two authors (CS and one of RM or KL) through a two-phase screening process. In the first phase, all abstracts will be screened by one author (CS). A 10\% random sample of the identified abstracts will be screened by a second author (either RM or KL) to identify those for inclusion. If there is disagreement over whether a study should be included between reviewers, then the study will move to the second phase of screening where the full text will be evaluated against the eligibility criteria by both reviewers and inclusion will be determined via discussion among the 3 reviewers. Studies that are found to have met all the inclusion and none of the exclusion criteria will be included.

\subsection{Data extraction and management}
Covidence will be used to extract and store the data from the review. A data extraction questionnaire was developed (Supplementary Material 1) and was piloted for use by CS and RM using a sample of 10 studies, with changes being made to the questionnaire where necessary. CS will extract data from all eligible studies in the review. Double data extraction will be performed by either RM or KL on a random sample of 10\% of eligible studies and additionally when there is uncertainty about studies to ensure consistency throughout the data extraction process. 

A full list of the data extraction items is provided in Table \ref{tab:table2}. We will only extract data that is reported in the article and supplementary material. We expect some data will be challenging to extract. The assumptions and simplifications that we will make under these conditions are summarised in Supplementary Table 1. If any post-hoc assumptions or simplifications are made, these will be reported as part of the analysis. 

\subsection{Analysis}
Once data extraction is complete, the data will be exported from Covidence. The extracted data will be cleaned and analysed using R \cite{R}. The data will be summarised using descriptive statistics. Frequencies and percentages will be reported to summarise categorical data. Medians and interquartile ranges will be used to summarise continuous data. The data and code will be made available publicly on Github.

\renewcommand{\arraystretch}{1.4}
\begin{table}[H]
\caption{Summary of items that were reported in the article that will be extracted}
\centering
\resizebox{\columnwidth}{!}{%
\begin{tabular}{|l|l|}
\hline
\textbf{Category} & \textbf{Extracted data} \\ \hline
\textbf{Study characteristics} & \begin{tabular}[c]{@{}l@{}}
\tabitem Title \\
\tabitem First author name(s) \\
\tabitem Publication year \\
\tabitem Funding source \\ 
\tabitem Journal \\
\tabitem Trial type \\
\end{tabular} \\ \hline
\textbf{Subject matter} & \begin{tabular}[c]{@{}l@{}}
\tabitem Setting \\
\tabitem Medical condition studied \\
\end{tabular} \\ \hline
\textbf{Design} & \begin{tabular}[c]{@{}l@{}}
\tabitem Ordinal outcome type \\
\tabitem Number of categories in the outcome\\
    \tabitem Whether the ordinal outcome was measured at a single time point or as a measure of change\\
    \tabitem Whether the categories of the ordinal scale were clearly defined, ordered, mutually exclusive\\ and, if a measure of change, symmetrical \\
    \tabitem Whether the ordinal outcome was a primary or secondary outcome \\
    \tabitem Whether sample size determination was used based off the ordinal outcome\\
    \tabitem Number of study participants included in the analysis (largest if multiple analyses)\\
\end{tabular} \\ \hline
\textbf{Statistical methods} & \begin{tabular}[c]{@{}l@{}}
\tabitem The statistical model(s) or method(s) that were used to analyse the ordinal outcome \\
\tabitem Type of inference used (frequentist/Bayesian) \\
\tabitem Target parameter used \\
\tabitem How the distribution of the ordinal outcome was summarised by intervention\\
\tabitem Methods used to account for clustering/repeated measures over time (if applicable) \\ 
\tabitem Details on whether the model assumptions were reported \\
\tabitem Whether the analysis that was reported in the results differed from the analysis\\ outlined in the methods section of the manuscript \end{tabular} \\ \hline
\textbf{Software included} & 
\tabitem Statistical software package used for the analysis \\ \hline
\end{tabular}%
}
\label{tab:table2}
\end{table}

\subsection{Patient and public involvement}
There will be no patients or public involvement in this review.

\subsection{Reporting}
The findings from this review will be reported using the Preferred Reporting Items for Systematic Reviews and Meta-Analyses extension for Scoping Reviews (PRISMA-ScR) checklist \cite{tricco2018prisma}. 

\section{DISCUSSION}
This paper describes a protocol for a scoping review that aims to examine the use, analysis and reporting of ordinal outcomes in RCTs in the top four medical journals. To our knowledge, there has not been a review of how ordinal outcomes are used in RCTs, particularly in the last decade. This review aims to address this gap and identify how ordinal outcomes are used in trials to improve our understanding of the appropriate analysis of such outcomes. 

\subsection{Strengths and limitations}
A targeted review of RCTs in top medical journals publishing original and recent research will highlight the current state of practice for analysing ordinal outcomes. We have \textit{a priori }specified eligibility criteria and strategies to handle anticipated challenges with data extraction. The screening and data extraction process will be conducted systematically, in which the pilot tests and double data extraction ensure the consistency and reliability of the extracted data. The search strategy, dataset and code will be made publicly available to ensure that the review is reproducible. The PRISMA-ScR checklist will be used to ensure that reporting is conducted to the highest standard. 

This review does, however, have its limitations. Although there is an exclusive focus on the PubMed database and on the top four medical journals, we made this decision given the scoping nature of the review, to make it as reproducible as possible, and to ensure that the total number of studies included in the review was manageable. We assume that these journals capture the current best practice in the use of ordinal outcomes in RCTs, and therefore reflect the current use of ordinal outcomes in practice. Our search strategy may miss certain phrases or variants (particularly related to an ordinal outcome), although our piloting phase has hopefully mitigated this to a large degree. We deliberately avoided including the names of specific scales in our search strategy as this would provide a non-representative sample. 

\subsection{Implications of this research}
In addition to critically appraising and examining the literature regarding the use of ordinal outcomes in RCTs, this review will identify areas of improvement for the use and the analysis of ordinal outcomes for future trials to ensure the reliability and transparency of reporting of such outcomes. We also hope the results will be used to inform methodological research in the analysis of ordinal outcomes.

\textbf{Funding sources/sponsors }  This work forms part of Chris Selman’s PhD, which is supported by an Australian Postgraduate Award administered through The University of Melbourne, Australia and a scholarship from the Australian Trials Methodology Research Network. Research at the Murdoch Children’s Research Institute is supported by the Victorian Government’s Operational Infrastructure Support Program. This work was supported by the Australian National Health and Medical Research Council (NHMRC) Centre for Research Excellence grants to the Victorian Centre for Biostatistics (ID1035261) and to the Australian Trials Methodology Research Network (ID1171422). Katherine Lee is funded by an NHMRC Career Development Fellowship (ID1127984).

\textbf{Authors' contributions }  CS, RM and KL conceived the study and CS wrote the first draft of the manuscript. All authors contributed to the design of the study, revision of the manuscript, and take public responsibility for its content.

\textbf{Competing interests statement }   None declared. 

\textbf{Acknowledgements } 
Chris Selman acknowledges the support of the University of Melbourne, Murdoch Children's Research Institute and the Australian Trials Methodology Research Network. Robert Mahar acknowledges the support of the Australian Trials Methodology Research Network Seed Funding Program. 

\textbf{Ethics and Dissemination }
As data and information will only be extracted from published studies, ethics approval will not be required. Conference presentations and peer-reviewed publications will be used to disseminate the results.

\newpage

% References
\bibliographystyle{aichej}
\bibliography{references}

\newpage 
\section{Supplementary Material}
\textbf{Supplementary Material 1: Data Extraction Questionnaire}

\textbf{Study Characteristics}

1.	What is the title of the manuscript?

2.	What is the name of the first author?

3.	What was the year of publication?

(1)	2012

\vspace{-0.6cm}
(2)	2013

\vspace{-0.6cm}
(3)	2014

\vspace{-0.6cm}
(4)	2015

\vspace{-0.6cm}
(5)	2016

\vspace{-0.6cm}
(6)	2017

\vspace{-0.6cm}
(7)	2018

\vspace{-0.6cm}
(8)	2019

\vspace{-0.6cm}
(9)	2020

\vspace{-0.6cm}
(10) 2021

\vspace{-0.6cm}
(11) 2022

4.	a) What was the funding source? (tick all that apply)

(1)	Non-profit

\vspace{-0.6cm}
(2)	Not reported

\vspace{-0.6cm}
(3)	Public

\vspace{-0.6cm}
(4) Industry

\vspace{-0.6cm}
(9) Other 

\vspace{-0.6cm}
(99) Unclear/unknown

4. b) If other, specify the other funding source.

5. What journal was the article published?

(1) BMJ

\vspace{-0.6cm}
(2) JAMA

\vspace{-0.6cm}
(3) The Lancet 

\vspace{-0.6cm}
(4) NEJM

6. Was an adaptive design used in the trial?

(0) No 

\vspace{-0.6cm}
(1) Yes 

\vspace{-0.6cm}
(99) Unknown/could not be determined

\textbf{Subject Matter}

1. Was the study conducted in a medical setting?

(1) Medical setting

\vspace{-0.6cm}
(2) Non-medical setting 

2. If the study was conducted in a medical setting, what medical condition was under study?

\textbf{Design} 

1. What type of ordinal scale was used?

(1) Single-state ordinal scale

\vspace{-0.6cm}
(2) Transition-state ordinal scale 

\vspace{-0.6cm}
(99) Unknown/could not be determined

2. Was the ordinal outcome a primary or secondary outcome?

(1) Primary outcome

\vspace{-0.6cm}
(2) Secondary outcome

3. How many categories did the ordinal outcome have? (record as an integer, N/A if unclear)

4. What properties of an ordinal scale did the ordinal outcome satisfy? (tick all that apply)

(1) Clearly defined and unambiguous categories 

\vspace{-0.6cm}
(2) Mutually exclusive categories 

\vspace{-0.6cm}
(3) Categories ordered in a hierarchical manner

\vspace{-0.6cm}
(4) The ordinal scale used can detect improvement and deterioration

\vspace{-0.6cm}
(5) Symmetrical scale (if the ordinal outcome is a transition-state scale)

\vspace{-0.6cm}
(99) Unknown/could not be determined

5. a) What did the ordinal outcome measure? (tick all that apply)

(1) Mortality/survival 

\vspace{-0.6cm}
(2) Clinical outcomes (e.g. (time to) treatment success/failure, severity scores, symptoms)

\vspace{-0.6cm}
(3) Physiological outcomes (e.g. viral detection/load, biomarkers)

\vspace{-0.6cm}
(4) Adverse events

\vspace{-0.6cm}
(5) Life impact (e.g. quality of life, compliance, mental health, satisfaction)

\vspace{-0.6cm}
(6) Resource use (e.g. economic, hospital, need for further intervention)

\vspace{-0.6cm}
(9) Other 

\vspace{-0.6cm}
(99) Unknown/could not be determined

5. b) If other, describe what the ordinal outcome measured.

6. a) Was the sample size determined in advance based on the ordinal outcome?

(0) No

\vspace{-0.6cm}
(1) Yes

\vspace{-0.6cm}
(99) Unknown/could not be determined

6. b) If yes, what methods did the authors use to determine the sample size?

(1) Analytical 

\vspace{-0.6cm}
(2) Simulation 

\vspace{-0.6cm}
(3) Ad-hoc

\vspace{-0.6cm}
(9) Other

\vspace{-0.6cm}
(99) Not applicable

7. What was the number of study participants included in the model used to analyse the ordinal outcome (largest N if multiple analyses)? (record as an integer)

\textbf{Statistical methods} 

1. a) How was the distribution of the ordinal outcome summarised by intervention group? (tick all that apply)

(1) Odds (category-specific)

\vspace{-0.6cm}
(2) Frequencies and proportions/percentages (category-specific)

\vspace{-0.6cm}
(3) Means (across all categories)

\vspace{-0.6cm}
(4) Medians (across all categories)

\vspace{-0.6cm}
(5) Standard deviations (across all categories)

\vspace{-0.6cm}
(6) Interquartile ranges (across all categories)

\vspace{-0.6cm}
(7) Summaries by group not used

\vspace{-0.6cm}
(9) Other 

\vspace{-0.6cm}
(999) Unknown/could not be determined

1. b) If other, describe how the distribution of the ordinal measure was summarised by intervention group.

2. If inferential statistics were used to analyse the ordinal outcome, what type of inference was used?

(1) Frequentist inference

\vspace{-0.6cm}
(2) Bayesian inference

\vspace{-0.6cm}
(99) Unknown/could not be determined

\vspace{-0.6cm}
(999) Not applicable

3. a) What was the reported target parameter? (tick all that apply)

(1) Odds ratio(s)

\vspace{-0.6cm}
(2) Difference in means

\vspace{-0.6cm}
(3) Difference in medians

\vspace{-0.6cm}
(4) Risk difference

\vspace{-0.6cm}
(5) Risk ratio

\vspace{-0.6cm}
(6) Descriptive only

\vspace{-0.6cm}
(7) Non-parametric procedure used

\vspace{-0.6cm}
(9) Other 

\vspace{-0.6cm}
(99) Unknown/could not be determined

3. b) If other, specify the reported target parameter.

4. a) Which statistical model(s) or method(s) did the authors report to analyse the ordinal outcome? (tick all that apply)

	(1) Cumulative logit model 
	
\vspace{-0.6cm}
	(2) Continuation ratio model 
	
	\vspace{-0.6cm}
	(3) Ordinal probit model 
	
	\vspace{-0.6cm}
	(4) Adjacent category logit model
	
\vspace{-0.6cm}
	(5) Logistic model 
	
\vspace{-0.6cm}
	(6) Linear model 
	
\vspace{-0.6cm}
	(7) Baseline category logit model
	
\vspace{-0.6cm}
	(8) Cox proportional hazards model

\vspace{-0.6cm}	
	(9) Mann-Whitney U-test
	
\vspace{-0.6cm}
	(10) Cochran-Mantel-Haenszel test

\vspace{-0.6cm}	
	(99) Other 
	
\vspace{-0.6cm}
    (999) None

\vspace{-0.6cm}
    (9999) Unknown/could not be determined

4. b) If other, specify the statistical model or method that was used to analyse the ordinal outcome.

5. a) Did the authors report in the manuscript whether a different statistical model or method had to be used in the analysis of the ordinal outcome? (e.g. the authors used a partial proportional odds model due to violation of the proportional odds assumption)

(0) No 

\vspace{-0.6cm}
(1) Yes

\vspace{-0.6cm}
(99) Unknown

\vspace{-0.6cm}
(999) Not applicable

5. b) If yes, provide detail as to which statistical model(s) or method(s) were in the initial analysis of the ordinal outcome and explain, if possible, why these were not used by the authors in the reported analysis.

6. a) Was the validity of the assumptions required for the reported statistical model/method(s) checked, justified and clearly described by the authors?

	(0) No 
	
\vspace{-0.6cm}
	(1) Yes 
	
\vspace{-0.6cm}
(99) Unknown/could not be determined

\vspace{-0.6cm}
(999) Not applicable

6. b) If yes, how did the authors check the assumptions of the statistical model/method? (tick all that apply)

(1) Graphical methods

\vspace{-0.6cm}	
	(2) Statistical methods
	
\vspace{-0.6cm}
    (3) Prediction methods

\vspace{-0.6cm}	
	(9) Other

\vspace{-0.6cm}
(99) Unknown/could not be determined
    
\vspace{-0.6cm}
    (999) Not applicable

6. c) If other, please describe.

7. a) If the ordinal outcome was measured repeatedly over time, how did the authors account for repeated measures in the analysis of the ordinal outcome? (tick all that apply)

	(1) Robust standard errors

\vspace{-0.6cm}	
	(2) Generalised estimating equations 

\vspace{-0.6cm}	
	(3) Mixed effects models 

\vspace{-0.6cm}	
	(4) Discrete-time Markov transition models 

\vspace{-0.6cm}	
	(5) Continuous-time Markov transition models 
	
\vspace{-0.6cm}
    (6) Adjusted for baseline measurement 

\vspace{-0.6cm}	
	(7) No methods used to account for repeated measures 

\vspace{-0.6cm}	
	(9) Other

\vspace{-0.6cm}
(99) Unknown/could not be determined

\vspace{-0.6cm}
    (999) Not applicable (i.e. only one measure was used in the analysis of the outcome)

7. b) If other, provide details.

\textbf{Software}

1. What statistical software package was used for the analysis? (tick all that apply)

	(1) R

\vspace{-0.6cm}	
	(2) SAS

\vspace{-0.6cm}	
	(3) SPSS

\vspace{-0.6cm}	
	(4) Stata

\vspace{-0.6cm}	
	(9) Other 

\vspace{-0.6cm}	
	(99) Unknown/not stated

1. b) If other, specify what software was used.

\textbf{Other}

1. Detail any other information that is deemed relevant and useful for the review.

\newpage
\renewcommand{\arraystretch}{1.4}
Supplementary Table 1: Anticipated challenges with data extraction and how they will be handled.
\begin{table}[H]
\centering
\resizebox{\columnwidth}{!}{%
\begin{tabular}{|l|l|l|}
\hline
\textbf{Challenge for data extraction} & \textbf{Category of items affected} & \textbf{How challenge will be handled} \\ \hline
\begin{tabular}[c]{@{}l@{}}Articles may have more than one \\ publication.\end{tabular} & Inclusion criteria & \begin{tabular}[c]{@{}l@{}}For articles with more than one publication date (such as \\ early-view/online publication or preliminary results) only\\ one publication date is required to be between 1 January\\ 2012 and 31 July 2022. The earlier date will be recorded\\ if two or more publication dates are between 1 January\\ 2012 and 31 July 2022. If there are two publications that\\ include a preliminary reporting of results followed by the\\ main reporting of results, then data from the main article\\ will be extracted.\end{tabular} \\ \hline
\begin{tabular}[c]{@{}l@{}}There are multiple trials in the\\ same manuscript.\end{tabular} & Inclusion criteria & \begin{tabular}[c]{@{}l@{}}If there are multiple trials included in the same manuscript but\\ are all sufficiently different from each other, all trials that \\use an ordinal outcome will be included in the review. If \\any trials are very similar to each other (e.g. share the same\\ protocol and/or statistical analysis plan), the first trial\\ that is mentioned and uses an ordinal outcome will be included\\ in the review. If the results of multiple trials are pooled,\\ then data will be extracted as if it were coming from a single trial. \end{tabular} \\ \hline
\begin{tabular}[c]{@{}l@{}}If a study uses more than one\\ statistical model to analyse \\the ordinal outcome.\end{tabular} & Data extraction (statistical methods) & \begin{tabular}[c]{@{}l@{}} All statistical models/methods will be recorded. For example,\\ it is possible that a study initially describes in the methods\\ section that a proportional odds model will be used. However,\\ if the proportional odds assumption was violated, the study\\ may use an alternative model to analyse the outcome in the\\ final analysis. We will only report statistical methods that \\were reported in the final analysis. \end{tabular} \\ \hline
\begin{tabular}[c]{@{}l@{}}There are multiple ordinal outcomes \\ that are used in the study.\end{tabular} & Any data extraction categories & \begin{tabular}[c]{@{}l@{}}If there are multiple ordinal outcomes, it is likely that the \\study  will use similar, if not the same, statistical methods to\\ analyse the outcome. Consequently, we will only examine how\\ the study has used the first ordinal outcome for the data\\ extraction process.\end{tabular} \\ \hline
\begin{tabular}[c]{@{}l@{}}There are two or more target parameters used\\ for the same ordinal scale.\end{tabular} & Any data extraction categories & \begin{tabular}[c]{@{}l@{}}It is reasonable to assume that a study would be interested\\ in potentially more than one target parameter using the\\ same ordinal scale if, for example, the study was investigating\\ both the proportion of absolute attainment of a pre-specified\\ outcome, and if there was an improvement in the distribution\\ of scores on the original ordinal scale. If an article has included\\ two or more target parameters using the same ordinal scale,\\ we will extract data only for the outcome that is analysed\\ on the original ordinal scale.\end{tabular} \\ \hline 
\end{tabular}
}
\label{tab:supptable2}
\end{table}

\end{document}